\RequirePackage{fix-cm}
\documentclass{svjour3}
\smartqed 
\usepackage{graphicx}
\usepackage{natbib}
\usepackage{amsmath}
\usepackage{hyperref}
\usepackage{multirow}

\begin{document}
\title{Extended two-body problem for rotating rigid bodies
}

\author{Alex Ho         \and
        Margrethe Wold \and 
        John T. Conway \and
        Mohammad Poursina
}

\institute{A. Ho \and M. Wold \and John. T. Conway \and M. Poursina  \at
              University of Agder, Jon Lilletuns vei 9 N-4879 Grimstad, Norway \\
              \email{alex.ho@uia.no}
}

\date{Received: date / Accepted: date}

\maketitle

\begin{abstract}
A new technique that utilizes surface integrals to find the force, torque and potential energy between two non-spherical, rigid bodies is presented. The method is relatively fast, and allows us to solve the full rigid two-body problem for pairs of spheroids and ellipsoids with 12 degrees of freedom. We demonstrate the method with two dimensionless test scenarios, one where tumbling motion develops, and one where the motion of the bodies resemble spinning tops. We also test the method on the asteroid binary (66391) 1999 KW4, where both components are modelled either as spheroids or ellipsoids. The two different shape models have negligible effects on the eccentricity and semi-major axis, but have a larger impact on the angular velocity along the $z$-direction. In all cases, energy and total angular momentum is conserved, and the simulation accuracy is kept at the machine accuracy level.
\keywords{Spheroids \and Dynamical evolution  \and Two rigid body problem \and Binary systems}
\end{abstract}
\section{Introduction}
\label{intro}
Asteroids are remnants from the formation of the solar system. It is hoped that detailed study of asteroids may improve our understanding on how the solar system was formed. One way to obtain a better knowledge of asteroids is to study the dynamics of multi-body systems. Roughly two decades ago, the existence of asteroid satellites was still uncertain \citep{Widenschilling_etal_1989}. Not long after, the first asteroid binary, (243) Ida and its satellite Dactyl, was discovered by the Galileo spacecraft mission in 1993 \citep{Chapman_etal_1995}. Since then, many more binaries have been observed and it is believed that roughly 15\% of near-Earth asteroids, larger than 200 meters in diameter, are binaries \citep{Margot_etal_2002}. A few asteroid binaries have also been observed amongst the main-belt asteroids, but the frequency of such systems is generally lower \citep{Merline_etal_2002}. 

Studying the dynamical evolution of two rigid bodies, including both their translational and rotational motion, is known as the full two-body problem. In the full two-body problem, the rotational and translational motion of both bodies are fully coupled \citep{1995CeMDA..63....1M}. 
However, studying the full two-body problem is not trivial for irregular bodies, such as asteroids, as no analytical solution exists to compute the gravitational force between two non-spherical objects. Therefore, several numerical approaches have been developed to solve the full-two body problem. 

A straightforward method of modelling the gravitational field of an asteroid is the mascon model \citep{Geissler_etal_1996}. In this approach, a set of point masses are placed in a grid, forming the shape of the asteroid. The total gravitational field of the asteroid will then be the sum of all the point mass fields. The advantage of the mascon model is that it can produce an accurate shape representation of an arbitrary body. Despite this advantage, the mascon model also has several drawbacks. For instance, the accuracy of the model depends on the number of point masses included in the body, and including a large number of point masses is computationally expensive. Furthermore \citet{Werner_Scheeres_1997} have shown that, despite including a large number of point masses, there exists significant errors in the force computation. This is due to the errors in the resolution of the surface of the asteroid, as the topology of the body is replaced with spherical balls. \citet{wittick2017mascon} have revisited and optimized the model to make it less computationally demanding and increased the numerical accuracy.

A different, yet common method to model the gravitational field of non-spherical bodies is through the application of series expansions. The most common case is generated by using expansions of spherical harmonics \citep{Konopliv2011}. However, for spherical harmonics, the potential can only be computed outside a given sphere, known as the Brillouin sphere, as the potential will diverge inside this region \citep{Moritz_1980}.  Spherical harmonics are also utilized in the full two-body problem to describe the mutual gravitational potential \citep{SCHEERES199667, 2016MNRAS.461.3982B, 2017CeMDA.128..261B}. 

Alternative parametrizations are also used to mitigate the divergence problem present in the spherical harmonics approach. For example, ellipsoidal harmonics has been used to compute the gravity field on various small solar system bodies \citep{2001CeMDA..79..235R, doi:10.1029/2001GL013768, 2002A&A...387.1114D, Reimond_Baur_2016}. However, the use of ellipsoidal harmonics can be cumbersome, due to the mathematical and numerical complexity. Alternatively, prolate spheroidal harmonics \citep{Fukushima_2014, SEBERA201670} provides simpler mathematical expressions compared to ellipsoidal harmonics, while providing a good geometric fit for non-spherical objects. Furthermore, \citet{Reimond_Baur_2016} have shown that, although ellipsoidal harmonics are far more accurate than spherical harmonics, there are virtually no difference between prolate spheroidal and ellipsoidal harmonics. Nevertheless, because all these methods use mathematical expansions, there exists a limit to the number of terms included in the model. Neglecting higher order terms, therefore, results in truncation errors.

Another way to expand potentials is to expand the inertia integrals. \citet{2009CeMDA.104..103S} used this approach to study the stability of two ellipsoids restricted in the plane. \citet{2017CeMDA.127..369H} present a fast method to compute the mutual potential, also using inertia integrals, between two arbitrary bodies using recurrence relations. \citet{2014CeMDA.119..313C} utilize the STF tensor formalism, which allows one to determine the coupling between spherical harmonics in a compact way, to determine the mutual potential and applied the method to study the (66391) 1999 KW4 binary system.

Another approach is to model an asteroid as a polyhedron of constant density.  Similar to the mascon model, the polyhedron model allows one to include finer geometric details of an asteroid. \citet{Werner_Scheeres_1997} present a method to compute the gravitational potential of a polyhedron and use it to model the gravity field of the asteroid (4769) Castalia. \citet{Conway_2015} gives an alternative formulation of the gravitational potential of a polyhedron through the use of vector potentials, a method used to study  how the dust is emitted and transported around Comet 67P \citep{Kramer_Noack_2015,Kramer_etal_2015}. For two bodies, a method was presented by \citet{2005CeMDA..91..337W} to determine the mutual potential between two polyhedra. This approach has been used to study the dynamical evolution of the asteroid binary (66391) 1999 KW4 \citep{2006Sci...314.1280S, 2006CeMDA..96..317F}. \citet{2017CeMDA.129..307S} present a different model for the mutual potential between a polyhedron and a rigid body of an arbitrary mass distribution and use this method to also study the (66391) 1999 KW4 system.

For most asteroids, detailed shape models are not available. An alternative to the polyhedron model is to approximate an asteroid with a well-defined shape, such as an ellipsoid. The gravitational potential of such bodies can be expressed analytically \citep{MacMillan1930}. However, the potential of an ellipsoid requires computation of elliptic functions, which may be computationally demanding. Nevertheless, using ellipsoid shape approximation to model asteroids has been e.g. used to study an ellipsoid-sphere system \citep{2004NYASA1017...81S, 2008CeMDA.100...63B}. 

A new approach, based on vector potentials, to compute the gravitational potential between two extended bodies is presented by \citet{Conway_2016}. Here, the force, torque and mutual potential energy are formulated as surface integrals under the assumption of constant density. This could potentially be a faster and more accurate technique, as the force and torque integrals are converted from volume integrals to surface integrals. Moreover, this approach does not rely on series expansions and will therefore not suffer from truncation errors. This approach was recently tested on coplanar spheroids and thin disks by \citet{2021CeMDA.133...27W}. In this paper, we extend the work of Wold \& Conway by giving a full three-dimensional treatment of the dynamics between the two bodies, including the coupling between translational and rotational motion.

In sec. \ref{sec:Math_framework}, we introduce the equations used to determine the translational and rotational motion and describe how we treat the rotational kinematics. In Sec. \ref{sec:Results_test}, we present the results of two dimensionless test scenarios, while in Sec. \ref{sec:Result_real} we apply our method on the asteroid binary system (66391) 1999 KW4. In this paper, all dotted variables denote the time derivative of the corresponding term. All vectors with a hat (e.g. $\hat{\mathbf{x}}$) show the vector expressions in the body-fixed frame (local frame), while those without a hat are in an inertial frame.

\section{Mathematical model}
\label{sec:Math_framework}
The main objective of the method is to calculate the force $\mathbf{F}$, torque $\mathbf{M}$ and potential $U$ of a solid body of uniform density $\rho $ in a gravitational field $\mathbf{g}\left(\mathbf{r}\right) $ in terms of surface integrals over the surface of the body. 

The force of an extended body with constant density $\rho$, due to an external gravitational field $\mathbf{g}(\mathbf{r})$, can be expressed through a volume integral
\begin{align}
\hat{\mathbf{F}} = \rho \iiint\limits_{V}\mathbf{g}(\mathbf{r})dV,
\end{align}
where $\mathbf{r}$ is a position vector to a point on the surface of the body being integrated over. If the gravitational field is given by the gradient of a scalar potential $\Phi(\mathbf{r})$, so that $\mathbf{g}(\mathbf{r}) = \nabla \Phi(\mathbf{r})$, we can apply the divergence theorem such that the force can be computed through a surface integral
\begin{align}
\hat{\mathbf{F}}=\rho \iint\limits_{S}\Phi( \mathbf{r}) \mathbf{n}dS.
\label{eq:Force_integral}
\end{align}
It can also be shown that the torque about the mass center of an extended body is given by
\begin{align}
\hat{\mathbf{M}}=-\rho \iint\limits_{S}\Phi ( \mathbf{r}) \mathbf{n}
\times \mathbf{r}dS  \label{eq:Torque_integral}
\end{align}
and the mutual potential as
\begin{align}
U=\frac{\rho }{3}\iint\limits_{S}\left( \mathbf{r}\Phi ( \mathbf{r}
) -\frac{1}{2}\left\vert \mathbf{r}\right\vert ^{2}\mathbf{g}\left( 
\mathbf{r}\right) \right) \cdot \mathbf{n}dS  \label{EqnA3_energy}
\end{align}
\citep{Conway_2016}. In these formulas $\mathbf{n}$ is a unit normal to the body surface. The above surface integrals are thus alternative expressions for the force, torque and potential between two bodies. In the following when we consider two interacting bodies, we replace the scalar potential $\Phi(\mathbf{r})$ with known analytical formulae for spheroids and ellipsoids as the first body, and integrate over either a spheroid or an ellipsoid as the second body. The surface integration method is outlined in detail by \citet{2021CeMDA.133...27W}.

Analytical formulas for $\mathbf{g}( \mathbf{r}) $ and $\Phi ( \mathbf{r}) $ in terms of elementary functions are given for spheroids and triaxial ellipsoids in \citep{MacMillan1930}. The Macmillan formulas for $\mathbf{g}( \mathbf{r}) $ were validated using Eq. \eqref{eq:Force_integral} by taking $\Phi ( \mathbf{r}) $ to be the scalar potential of a unit point mass and integrating this over a spheroid. Exact agreement was obtained between Eq. \eqref{eq:Force_integral} and the MacMillan formulas. The analytical expressions for the moments of inertia of spheroids and ellipsoids are already available. Therefore, it is straightforward to set up surface integration schemes in a local body-fixed coordinate system for spheroids and ellipsoids.

Here, we analyze the motion of the bodies in a global inertial coordinate system using the forces and torques acting on each body and applying Newton's laws. As we consider a two-body problem, if $\mathbf{F}_{A}$ and $\mathbf{F}_{B}$ are the global forces acting on the two bodies $A$ and $B$, then from Newton's third law
\begin{align}
\mathbf{F}_{B}=-\mathbf{F}_{A}\text{.}  \label{EqnA5}
\end{align}
It should be noted that $\hat{\mathbf{M}}_A$, which is the torque of $\hat{\mathbf{F}}_A$ about the origin of the body-fixed frame of body $A$, may not necessarily be equal to $\hat{\mathbf{M}}_B$, which is the torque of $\hat{\mathbf{F}}_B$ about the origin of the body-fixed frame of body $B$. However, the torque of all forces about any arbitrary predefined point in the system must be zero in order to maintain the conservation of the angular momentum.

When the forces and torques are computed, the equations of motion can be solved as a standard initial value problem, where the velocities $\mathbf{v}$ and positions $\mathbf{r}$ can be integrated as
\begin{align}
\frac{d\mathbf{v}}{dt} &= \frac{\mathbf{F}}{m} \\
\frac{d\mathbf{r}}{\mathrm{d}t} &= \mathbf{v}
\end{align}
and the angular velocity $\hat{\boldsymbol{\omega}}$ integrated from
\begin{align}
\frac{d\mathbf{J}}{dt} = \mathbf{M},
\end{align}
where $\mathbf{J}$ is the angular momentum of the body.

We use the embedded Runge-Kutta method of order 9(8) by Verner \citep{Verner2010} to solve the equations of motion. The coefficients of the Butcher tableau are taken from Verner's web page \footnote{\url{http://people.math.sfu.ca/~jverner/}}, using the `most efficient' coefficients provided in the web page. 

\subsection{Rotation angles and rotational motion}
\label{sec:RotAngles}
The force and torque integrals given in Eqs. \eqref{eq:Force_integral} and \eqref{eq:Torque_integral} are computed in the body-fixed frames. However, because we are interested in the motion in the inertial frame, the equations of motion must be projected back to the inertial frame. To move between different reference frames, we use a rotation matrix. In this paper, we adopt the Tait-Bryan convention where the rotation matrix $\mathcal{R}$ takes the form
\begin{align}
    &\mathcal{R}_x = \begin{bmatrix}
    1 & 0 & 0 \\
    0 & \cos\phi & -\sin\phi\\
    0 & \sin\phi & \cos\phi
    \end{bmatrix}\\
    &\mathcal{R}_y = \begin{bmatrix}
    \cos\theta & 0 & \sin\theta \\
    0 & 1 & 0\\
    -\sin\theta & 0 & \cos\theta
    \end{bmatrix}\\
    &\mathcal{R}_z = \begin{bmatrix}
    \cos\psi & -\sin\psi & 0 \\
    \sin\psi & \cos\psi & 0\\
    0 & 0 & 1
    \end{bmatrix}\\
    &\mathcal{R}(\phi,\theta,\psi) = \mathcal{R}_z\mathcal{R}_y\mathcal{R}_x.
    \label{eq:RotationMatrix}
\end{align}
The angles $(\phi, \theta, \psi)$ are the Tait-Bryan angles, and correspond to rotations around the body-fixed $x,y,z$ axes respectively\footnote{These angles are also known as roll, pitch and yaw.}. The rotation matrix given in Eq. \eqref{eq:RotationMatrix} is then used to project the force in the body-fixed frame, from Eq. \eqref{eq:Force_integral}, back to the inertial frame using the equation
\begin{align}
\mathbf{F} = \mathcal{R} \hat{\mathbf{F}}.
\label{eq:Force_global}
\end{align}
The computed torque in Eq. \eqref{eq:Torque_integral} is used to determine how the angular velocity changes over time, using the equations of motion for each body in its body-fixed frame, given as 
\begin{align}
    &I_{11}\dot{\hat{\omega}}_x + (I_{33} - I_{22})\hat{\omega}_y\hat{\omega}_z = \hat{M}_x  \label{eq:DiffEq_phidott} \\
    &I_{22}\dot{\hat{\omega}}_y + (I_{11} - I_{33})\hat{\omega}_x\hat{\omega}_z = \hat{M}_y \label{eq:DiffEq_thetadott} \\
    &I_{33}\dot{\hat{\omega}}_z + (I_{22} - I_{11})\hat{\omega}_x\hat{\omega}_y = \hat{M}_z
    \label{eq:DiffEq_psidott}
\end{align}
\citep{curtis2013orbital}, where $(\hat{\omega}_x, \hat{\omega}_y, \hat{\omega}_z)$ are the angular velocity components in the body-fixed frame, $I$ is the moment of inertia tensor and $(\hat{M}_x, \hat{M}_y, \hat{M}_z)$ are the components of the torque computed in the body-fixed frame. The angular velocity $(\hat{\omega}_x, \hat{\omega}_y, \hat{\omega}_z)$ is related to the angular velocity in the inertial frame as
\begin{align}
    \begin{bmatrix}
    \omega_x\\
    \omega_y\\
    \omega_z
    \end{bmatrix}= \mathcal{R}
    \begin{bmatrix}
    \hat{\omega}_x\\
    \hat{\omega}_y\\
    \hat{\omega}_z
    \end{bmatrix}.
\end{align}
The rotation angles ($\phi, \theta, \psi$) change over time, and the following kinematic equations relate the time rate of change of these angles to the angular velocity of the body
\begin{align}
\dot{\phi} &= \hat{\omega}_x + (\hat{\omega}_y\sin\phi + \hat{\omega}_z\cos\phi)\tan\theta \label{eq:DiffEq_phi}\\
\dot{\theta} &= \hat{\omega}_y\cos\phi - \hat{\omega}_z\sin\phi \label{eq:DiffEq_theta}\\
\dot{\psi} &= (\hat{\omega}_y\sin\phi + \hat{\omega}_z\cos\phi)\sec\theta\label{eq:DiffEq_psi}
\end{align}
\citep{fossen2011handbook}. These kinematic equations become singular when $\theta = n\pi/2$ for any odd integer $n$. This singularity is a mathematical problem and is normally resolved by using Euler parameters \citep[see e.g.][]{kane1983spacecraft}. Nevertheless, this singularity is generally not a problem for our cases in question.

\subsection{Surface integration}
The expressions for the surface elements in the surface integrals given by Eqs. \eqref{eq:Force_integral}-\eqref{EqnA3_energy}, as well as the components of the moment of inertia tensor, vary depending on the shape of the bodies. This can be generalized to any arbitrary ellipsoidal shapes. 

Consider a general ellipsoid with semiaxes $(a,b,c)$. The surface elements $\mathbf{n}dS$ and $\mathbf{n}\times \mathbf{r}dS$, used in Eqs. \eqref{eq:Force_integral}-\eqref{EqnA3_energy} respectively, for a general ellipsoid, are given as
\begin{align}
    \mathbf{n}dS &= 
    \begin{bmatrix}
    \frac{b\sqrt{c^2-z^2}}{c}\cos\alpha \\
    \frac{a\sqrt{c^2-z^2}}{c}\sin\alpha \\
    \frac{ab}{c^2}z
    \end{bmatrix}
    \mathrm{d}\alpha\mathrm{d}z  
    \label{eq:Force_element} \\
    \mathbf{n}\times\mathbf{r}dS &= 
    \begin{bmatrix}
    \frac{a}{c^3}(c^2-b^2)\sqrt{c^2-z^2}z\sin\alpha \\
    -\frac{b}{c^3}(c^2-a^2)\sqrt{c^2-z^2}z\cos\alpha \\
    \frac{b^2-a^2}{c^2}(c^2-z^2)\cos\alpha\sin\alpha
    \end{bmatrix}
    \mathrm{d}\alpha\mathrm{d}z
    \label{eq:Torq_element}
\end{align}
\citep{2021CeMDA.133...27W}, where $\alpha$ is the angle of the cylindrical coordinates (written as $\phi$ in Wold \& Conway) and $z$ is the generalized latitude line on the ellipsoid.
For a general ellipsoid with density $\rho$, the only non-zero components of the moment of inertia tensor $I_{ij}$ can be expressed analytically as
\begin{align}
    I_{11} &= \frac{4\pi \rho a b c}{15}(b^2+c^2) \\
    I_{22} &= \frac{4\pi \rho a b c}{15}(c^2+a^2) \\
    I_{33} &= \frac{4\pi \rho a b c}{15}(a^2+b^2).
\end{align}

\subsubsection{Rotated gravitational potential}
\begin{figure}
\centering
\includegraphics[width=\linewidth]{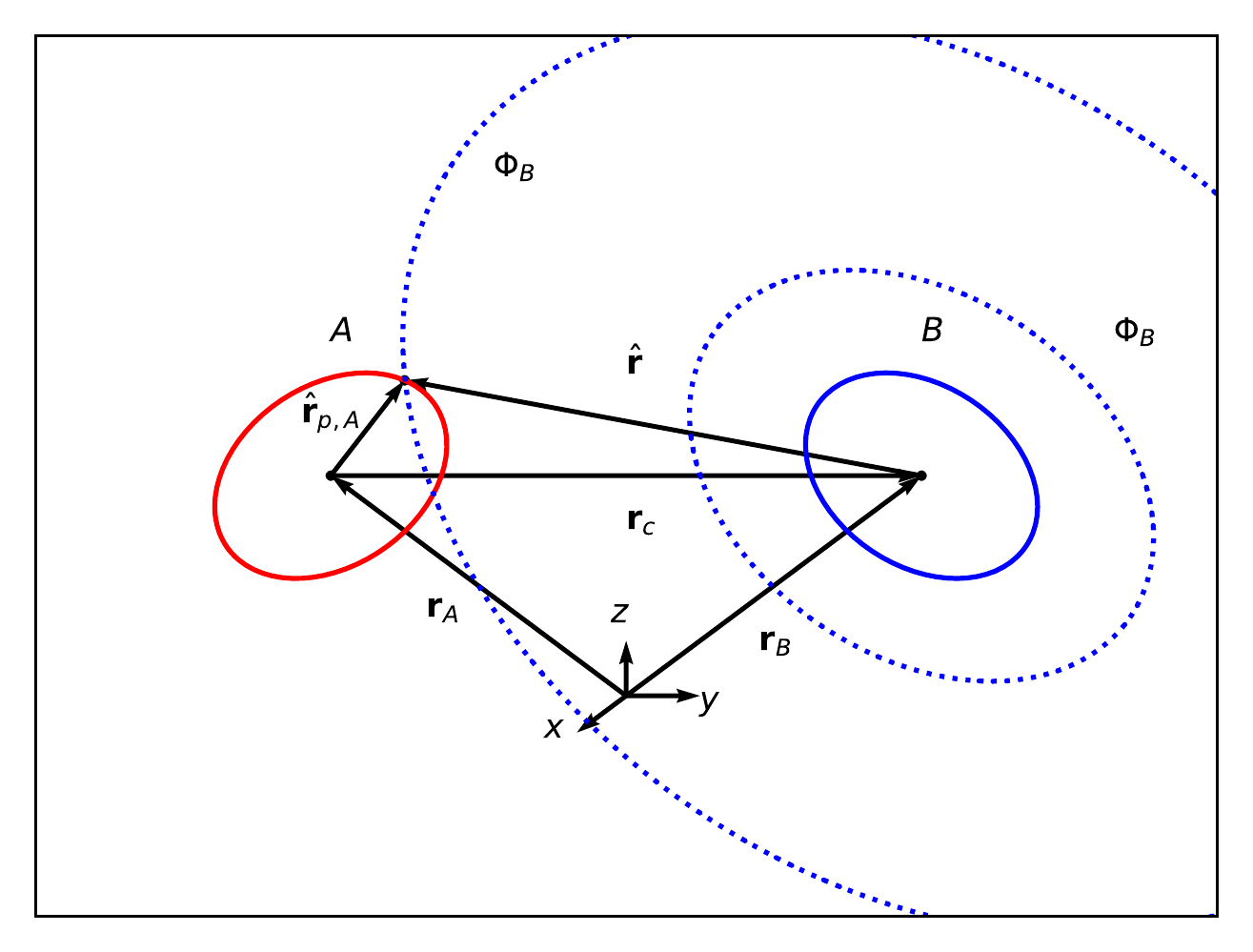}
\caption{An illustration on how the gravitational field is computed after the bodies have rotated. The gravitational potential $\Phi_B$ (blue dotted lines) is rotated according to the rotation of spheroid $B$. The axes $x,y,z$ denotes the inertial frame of the system.}
\label{fig:Rotated_field}
\end{figure}
In this section, we will describe how to incorporate the change in the gravitational potential $\Phi(\mathbf{r})$ and gravitational field $\mathbf{g}(\mathbf{r})$ when the bodies are rotated.

Consider two bodies $A$ and $B$ whose mass centers are respectively positioned by $\mathbf{r}_A$ and $\mathbf{r}_B$ from the origin of the inertial reference frame and with their respective rotation matrices $\mathcal{R}_A$ and $\mathcal{R}_B$. The vector linking the two centroids therefore becomes
\begin{align}
\mathbf{r}_c = \mathbf{r}_B - \mathbf{r}_A.
\end{align}
Let us consider the case where we compute the force on body $A$. In the inertial frame, the position vector of an arbitrary point located on the surface of $A$ is expressed as
\begin{align}
\mathbf{r}_p = \mathcal{R}_A\hat{\mathbf{r}}_{p,A} + \mathbf{r}_A,
\end{align}
where $\hat{\mathbf{r}}_{p,A}$ is a vector describing the surface points of $A$ in its body-fixed frame. For a general ellipsoid, this becomes
\begin{align}
\hat{\mathbf{r}}_{p,A} = 
\begin{bmatrix}
\frac{a}{c}\sqrt{c^2-z^2}\cos\alpha \\
\frac{b}{c}\sqrt{c^2-z^2}\sin\alpha \\
z
\end{bmatrix}.
\end{align}
The term $\mathbf{r}$ used as input argument in the $\Phi(\mathbf{r})$ and $\mathbf{g}(\mathbf{r})$ functions is a vector that points from the centroid of $B$ to any point on the surface of $A$. This can be written as
\begin{align}
\mathbf{r} = \mathbf{r}_p - \mathbf{r}_B = \mathcal{R}_A\hat{\mathbf{r}}_{p,a} - \mathbf{r}_c,
\label{eq:rprime_global}
\end{align}
which can be expressed in the body frame of reference as
\begin{align}
\hat{\mathbf{r}} = \mathcal{R}_B^T \mathbf{r} = \mathcal{R}_B^T(\mathcal{R}_A \hat{\mathbf{r}}_{p,A} - \mathbf{r}_c).
\label{eq:Rotated_local_point}
\end{align}
Equation \eqref{eq:Rotated_local_point} is then used as the input argument of both the gravitational potential $\Phi(\mathbf{r})$ and the gravitational field $\mathbf{g}(\mathbf{r})$. As the bodies rotate, so must the relative gravitational field $\mathbf{g}(\mathbf{r})$. This can be achieved using the following equation
\begin{align}
    \mathbf{g}_A(\hat{\mathbf{r}}) = \mathcal{R}^T_A \mathcal{R}_B \mathbf{g}_B(\hat{\mathbf{r}}).
\end{align}

\subsection{Total energy}
When all the differential equations are solved, we use the computed velocities, positions, angular velocities and rotation angles to determine the energy of the system. The translational kinetic energy of each body is computed as
\begin{align}
E_{k} = \frac{1}{2} m\mathbf{v}^2,
\end{align}
where $m$ and $\mathbf{v}$ are the mass and the center of mass velocity of the body, respectively. For an ellipsoid with constant density, the kinetic energy of rotational motion is
\begin{align}
E_{\text{rot}} = \frac{1}{2}\left(I_{11}\hat{\omega}_{x}^2 + I_{22}\hat{\omega}_{y}^2 + I_{33}\hat{\omega}_{z}^2\right).
\end{align}
The mutual potential energy $U$ is computed by Eq. \eqref{EqnA3_energy}.
The total energy of the system is the sum of the kinetic, rotational and potential energies. 
\subsection{Angular momentum}
\label{sec:Angular_momentum}
Because we do not include external forces and moments to model the system, the angular momentum and total energy of the system must be conserved. The total angular momentum, in the inertial frame, is 
\begin{align}
    \mathbf{J}_{tot} = \mathbf{J}_A + \mathbf{J}_B,
\end{align}
where the angular momentum of each body is given by
\begin{align}
    \mathbf{J} = \mathbf{r}\times m\mathbf{v} + \mathcal{R} I \boldsymbol{\hat{\omega}}.
\end{align}
As previously stated, for a general ellipsoid, the only non-zero components of the moment of inertia are $I_{11},I_{22}$ and $I_{33}$. Therefore, the total angular momentum becomes
\begin{align}
    \mathbf{J}_{tot} = \mathbf{r}\times m\mathbf{v} + \mathcal{R} \begin{bmatrix}
    I_{11} \hat{\omega}_{x} \\
    I_{22} \hat{\omega}_{y} \\
    I_{33} \hat{\omega}_{z}
    \end{bmatrix}.
    \label{eq:Total_angmom}
\end{align}
\subsection{Gravitational potential}
\label{sec:Grav_pot_spheroid}
To compute the force and torque given in Eqs. \eqref{eq:Force_integral} and \eqref{eq:Torque_integral}, we require an expression for the gravitational potential $\Phi(\mathbf{r})$. This potential can take different forms depending on the shape of the body. In this paper, we consider spheroids and triaxial ellipsoids, for which analytical expressions for $\Phi(\mathbf{r})$ are available.

For a general ellipsoid with semiaxes $a > b > c$ and constant density $\rho$, the gravitational potential is given by
\begin{align}
\Phi(\mathbf{r}) &= \frac{2\pi\rho abc}{\sqrt{a^2-c^2}}\Bigg(\left[1-\frac{x^2}{a^2-b^2} + \frac{y^2}{a^2-b^2}\right]F(\omega_\kappa, k) \nonumber \\
&+ \left[\frac{x^2}{a^2-b^2} - \frac{(a^2-c^2)y^2}{(a^2-b^2)(b^2-c^2)} + \frac{z^2}{b^2-c^2}\right]E(\omega_\kappa, k) \label{eq:Potential_elliptic}\\
&+ \left[\frac{c^2+\kappa}{b^2-c^2}y^2 - \frac{b^2+\kappa}{b^2-c^2}z^2\right]\frac{\sqrt{a^2-c^2}}{\sqrt{(a^2+\kappa)(b^2+\kappa)(c^2+\kappa)}}\Bigg) \nonumber
\end{align}
\citep{MacMillan1930}, where $F(\omega_\kappa,k)$ and $E(\omega_\kappa,k)$ are the elliptic integrals of the first and second kind respectively, $\kappa$ is the largest root of the equation
\begin{align}
\frac{x^2}{a^2+\kappa} + \frac{y^2}{b^2+\kappa} + \frac{z^2}{c^2+\kappa} = 1
\label{eq:kappa_eq}
\end{align}
and
\begin{align}
\omega_\kappa &= \sin^{-1}\sqrt{\frac{a^2-c^2}{a^2+\kappa}}
\label{eq:Ellipsepotential_omega}\\
k &= \sqrt{\frac{a^2-b^2}{a^2-c^2}}.
\label{eq:Ellipsepotential_K} 
\end{align}
For an oblate spheroid with semiaxes $a = b > c$, the gravitational potential can be expressed as
\begin{align}
    \Phi(\mathbf{r}) &= \frac{2\pi \rho a^2 c}{\sqrt{a^2 - c^2}}\left(1 - \frac{x^2 + y^2 - 2z^2}{2(a^2 - c^2)}\right)\times \sin^{-1}\sqrt{\frac{a^2-c^2}{a^2+\kappa}} \nonumber \\
    &+ \frac{\pi \rho a^2 c\sqrt{c^2+\kappa}}{a^2 - c^2}\frac{x^2+y^2}{a^2+\kappa} - \frac{\pi \rho a^2 c}{a^2-c^2}\frac{2z^2}{\sqrt{c^2+\kappa}}
    \label{eq:Oblate_spheroid}
\end{align}
\citep{MacMillan1930}, where $\kappa$ still satisfies Eq. \eqref{eq:kappa_eq}. It should be noted that the gravitational potentials given in Eqs. \eqref{eq:Potential_elliptic} and \eqref{eq:Oblate_spheroid} are exterior potentials. The expression for the gravitational field $\mathbf{g}(\mathbf{r})$, which is required to compute the mutual potential energy in Eq. \eqref{EqnA3_energy}, is derived in Appendix \ref{sec:Appendix_gravfield}.
  
\section{Dimensionless test scenarios}
\label{sec:Results_test}
\begin{table}
\caption{The initial parameters used for the two dimensionless test cases. The second and third columns show initial positions $\mathbf{r}_0$ and velocities $\mathbf{v}_0$ in the inertial frame, respectively. The fourth column indicates the initial rotation angles in radians. The fifth column shows the initial angular velocities in the body-fixed frames, in units of radians per unit time. The sixth and seventh columns show the semiaxes $(a,b,c)$ of the spheroids and their masses, respectively. The top and bottom rows, for each case, correspond to body $A$ and $B$ respectively.}
\label{table:Init_conditions}
\renewcommand{\arraystretch}{1.5}
\begin{tabular}{l|cccccc}
\hline\noalign{\smallskip}
Scenario & $\mathbf{r}_0$ & $\mathbf{v}_0$ & ($\phi_0, \theta_0, \psi_0$) & $\hat{\boldsymbol{\omega}}_0$ & ($a,b,c$) & $m$ \\
\noalign{\smallskip}\hline
Case 1 & $(-4,0,0)$ & $(0,0.3,0)$ & $(\frac{\pi}{32},0,0)$ & $(0,0,0)$ & $(1,1,0.25)$ & $2$ \\
(Rotated)& $(4,0,0)$ & $(0,-0.3,0)$ & $(-\frac{\pi}{32},0,0)$ & $(0,0,0)$ & $(1,1,0.25)$ & $2$ \\
\noalign{\smallskip}\hline
\multirow{2}{*}{Case 2 }& $(-4,0,0)$ & $(0,0.3,0)$ & $(0,\frac{\pi}{32},0)$ & $(0,0,2)$ & $(1,1,0.25)$ & $2$ \\
(Spinning) & $(4,0,0)$ & $(0,-0.3,0)$ & $(0,-\frac{\pi}{32},0)$ & $(0,0,2)$ & $(1,1,0.25)$ & $2$ \\
\noalign{\smallskip}\hline
\end{tabular}
\end{table}

We first test our method for two dimensionless cases, where the gravitational constant is set to unity, i.e. $G = 1$. In these scenarios, we let both bodies take spheroidal shapes. The initial conditions and spheroid parameters for the simulations are summarized in Tab. \ref{table:Init_conditions}. The time span of the simulations is $t\in[0,200]$. The actual simulation times on an ordinary laptop computer for these test scenarios varied from 20 seconds to two minutes.

\subsection{Case 1: Small initial rotation of the spheroids}
\label{sec:tumbling_system}
We first consider a case where the spheroids have a small initial rotation angle $\phi_{0,A} = - \phi_{0,B} = \pi/32$ about the $x$-axis and zero angular velocity. Their physical parameters and initial conditions are listed in Tab. \ref{table:Init_conditions} under ``Case 1 (Rotated)''. 

\begin{figure*}
\centering
\includegraphics[width=\linewidth]{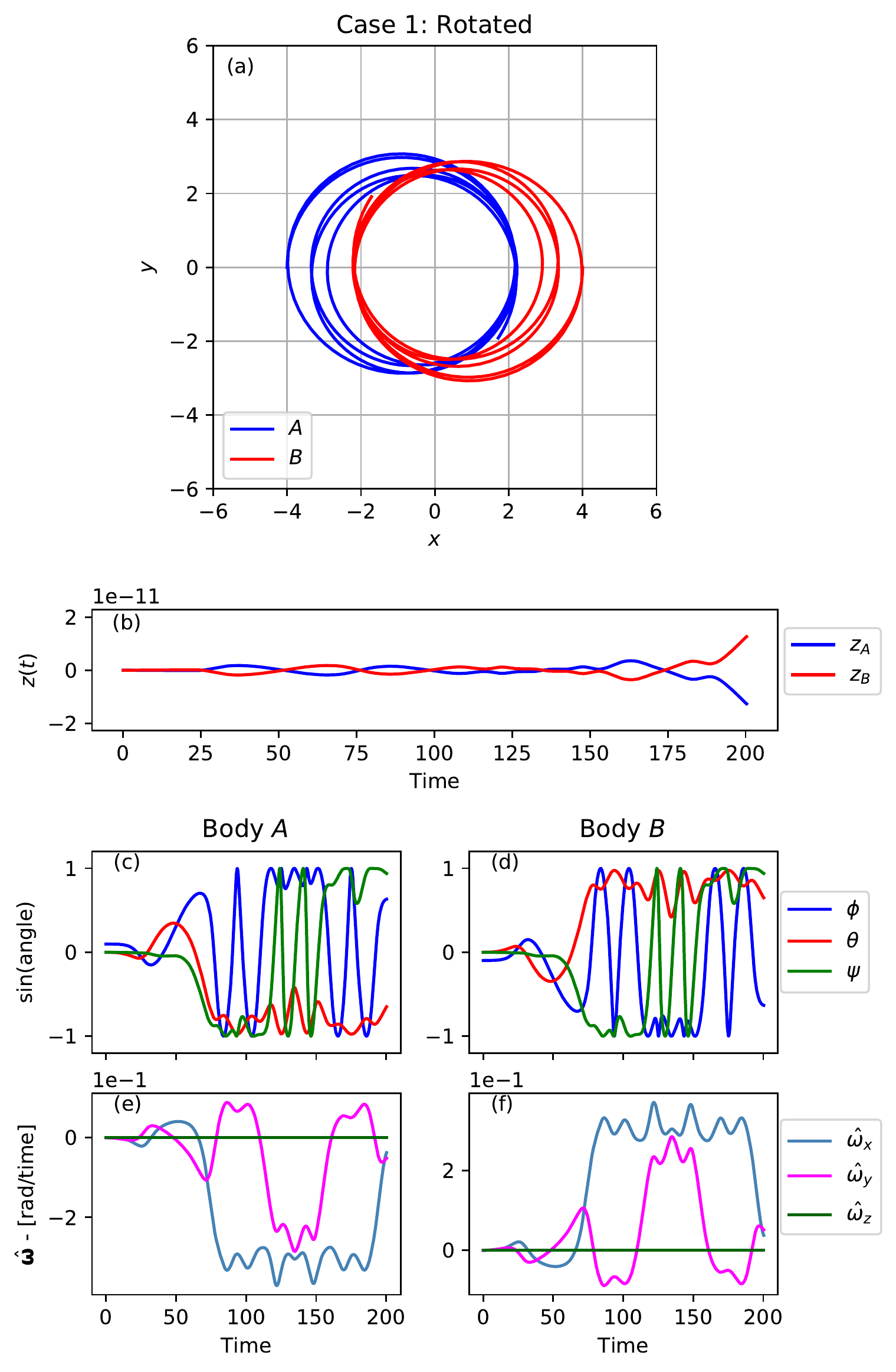}
\caption{For case 1, where the spheroids are initially rotated about their respective $x$-axes. \textbf{(a)} The orbits of spheroids $A$ and $B$ in the inertial frame projected into the $xy$-plane. \textbf{(b)} The $z$-position of the spheroids as a function of time. \textbf{(c,d)} The rotation angles of bodies $A$ and $B$ as sine functions. \textbf{(e,f)} The angular velocity in the body-fixed frames of bodies $A$ and $B$ as a function of time. The unit of the angular velocity $\boldsymbol{\hat{\omega}}$ is radians per unit time.}
\label{fig:Case2Resultsxrot}
\end{figure*}

\begin{figure}
\centering
\includegraphics[width=\linewidth]{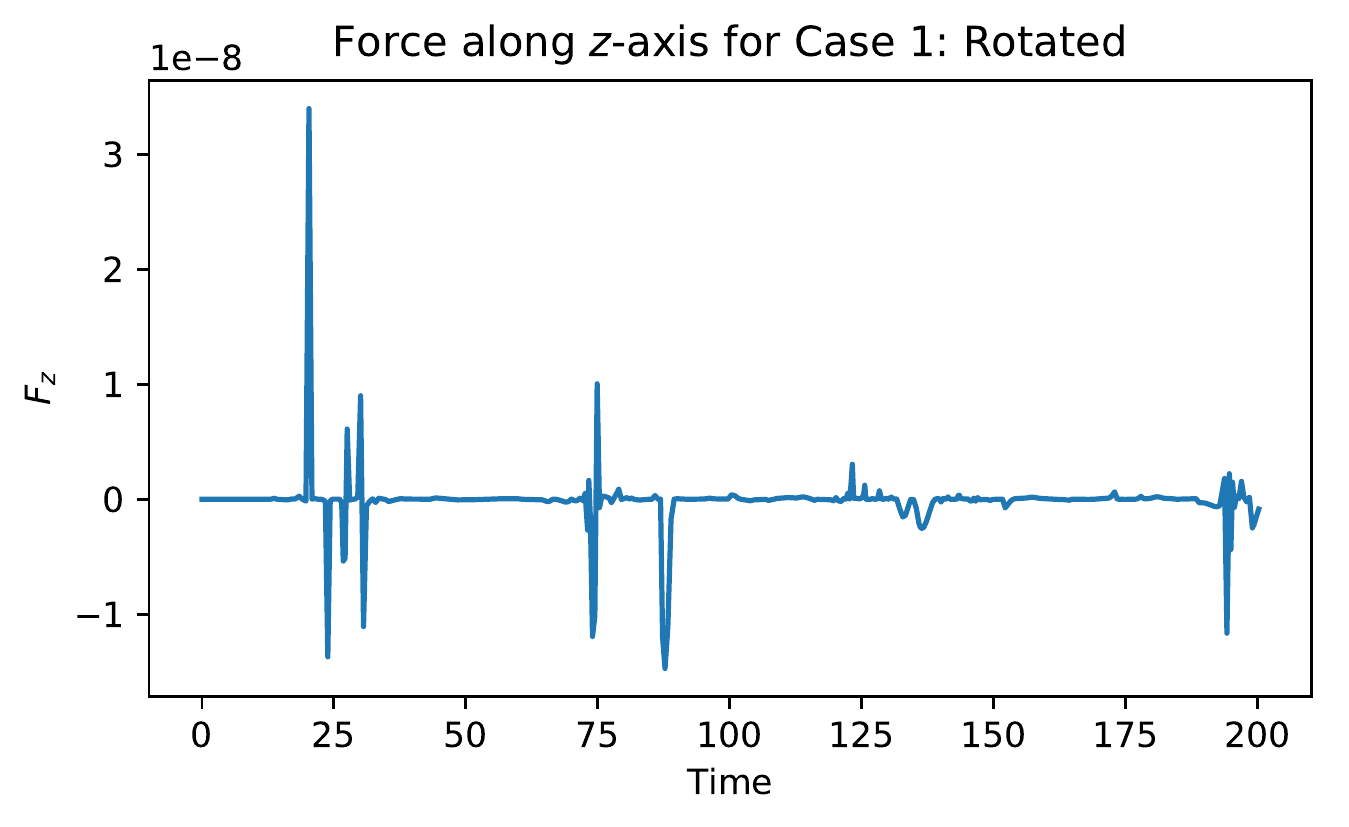}
\caption{$F_z$ component on body $A$ for case 1. The $F_x$ and $F_y$ components are not included in the figure as they take values of order $10^{-1}$.}
\label{fig:Case2Force}
\end{figure}

Fig. \ref{fig:Case2Resultsxrot}a shows the spheroid orbits projected into the $xy$-plane, while Fig. \ref{fig:Case2Resultsxrot}b shows their $z$-positions as functions of time. Because forces are acting parallel to the $z$-axis, as seen in Fig. \ref{fig:Case2Force}, the motions of the spheroids are no longer restricted to one plane. The motion along the $z$-axis is 11 orders of magnitude smaller than the motions along the $x$ and $y$-axes. The motion in the $z$-direction grows with simulation time, and becomes of the order $10^{-1}$ at $t \approx 2200$. The motion along the $x$ and $y$-directions, unlike the motion in $z$-direction, did not have any significant changes in the same time duration.

The rotation angles $(\phi,\theta,\psi)$, plotted as sine functions, for spheroids $A$ and $B$, are shown in Fig. \ref{fig:Case2Resultsxrot}c and Fig. \ref{fig:Case2Resultsxrot}d, respectively.
Both spheroids rotate multiple times about their $x$ and $z$-axes, and therefore have tumbling motion. The angular velocities, projected onto the respective body-fixed frames, are shown in Fig. \ref{fig:Case2Resultsxrot}e and \ref{fig:Case2Resultsxrot}f for spheroids $A$ and $B$. Due to the rotational symmetry of the bodies about their $z$-axes, no gravitational forces can change the rotational speeds about the $z$-axes. Hence, the body-fixed component $\hat{\omega}_z$ remains constant for both spheroids. 

\begin{figure}
\centering
\includegraphics[width=\linewidth]{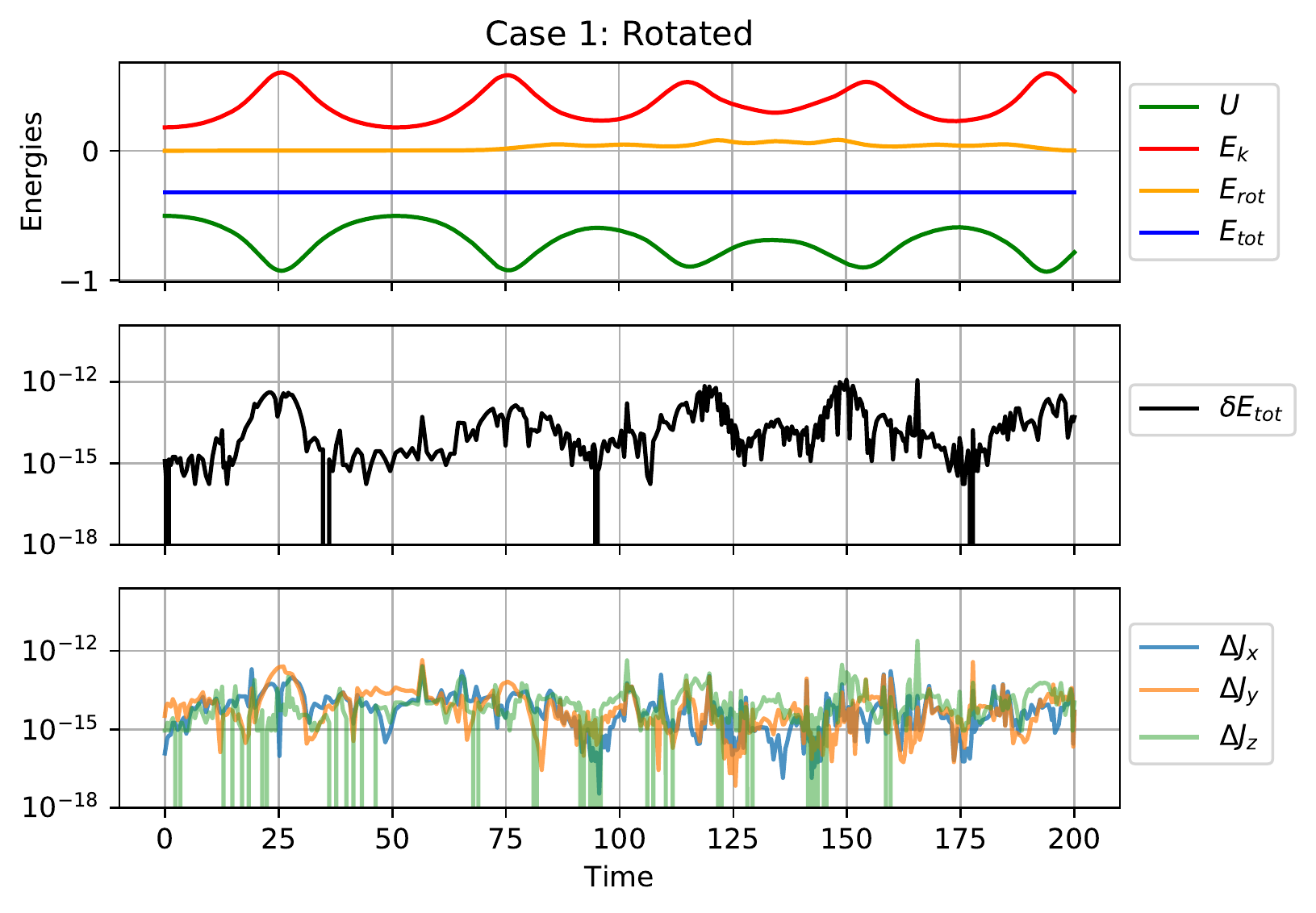}
\caption{For case 1. \textit{Top}: The potential energy $U$ (green), translational kinetic energy $E_k$ (red), rotational kinetic energy $E_{rot}$ (orange) and the total energy $E_{tot}$ (blue). \textit{Middle}: The relative error of the total energy. \textit{Bottom}: The absolute errors in the components of the total angular momentum. 
The errors in both the total energy $E_{tot}$ as well as the total angular momentum $\mathbf{J}_{tot}$ are smaller than $10^{-12}$, which demonstrates that these quantities are conserved in our simulations.}
\label{fig:Case2Energyxrot}
\end{figure}

The top panel of Fig. \ref{fig:Case2Energyxrot} shows the energies in the system. While both spheroids have no initial rotational motion, the rotational energy starts to make a noticeable contribution near $t\approx 75$, which is consistent with Figs. \ref{fig:Case2Resultsxrot}e and \ref{fig:Case2Resultsxrot}f, indicating that the angular velocities start to increase around $t=75$.
The middle panel of Fig. \ref{fig:Case2Energyxrot} shows the relative error in the total energy as a function of time. The relative error is computed as
\begin{align}
\delta E = \left|\frac{E_{i+1} - E_{i}}{E_{i}}\right|, \quad i = 0,1,2,...,(N_s-1)
\end{align}
where $N_s$ is the number of data points. 
Because the relative error is smaller than $10^{-12}$, the total energy can be considered as conserved in our simulation. 
Finally, the absolute error of the three total angular momentum components are shown in the bottom panel of Fig. \ref{fig:Case2Energyxrot}. The absolute error is computed as 
\begin{align}
\delta J = |J_{i+1} - J_{i}|, \quad i = 0,1,2,...,(N_s-1).
\end{align}
Similar to the total energy, the error in the total angular momentum is smaller than $10^{-12}$, and we conclude that also the total angular momentum is conserved in the simulation.

\subsection{Case 2: Adding initial angular velocity to the spheroids}
\label{sec:Spinning_system}
We now consider a system where both spheroids are initially spinning around their $z$-axes, with $\hat{\omega}_{0,z} = 2$, while the other angular velocity components are initially set to zero. We also let both spheroids start with a slight tilt, in which $\theta_{0,A} = -\theta_{0,B} = \pi/32$, while the other angles are initially zero. The other parameters are the same as those in case 1.

\begin{figure*}
\centering
\includegraphics[width=\linewidth]{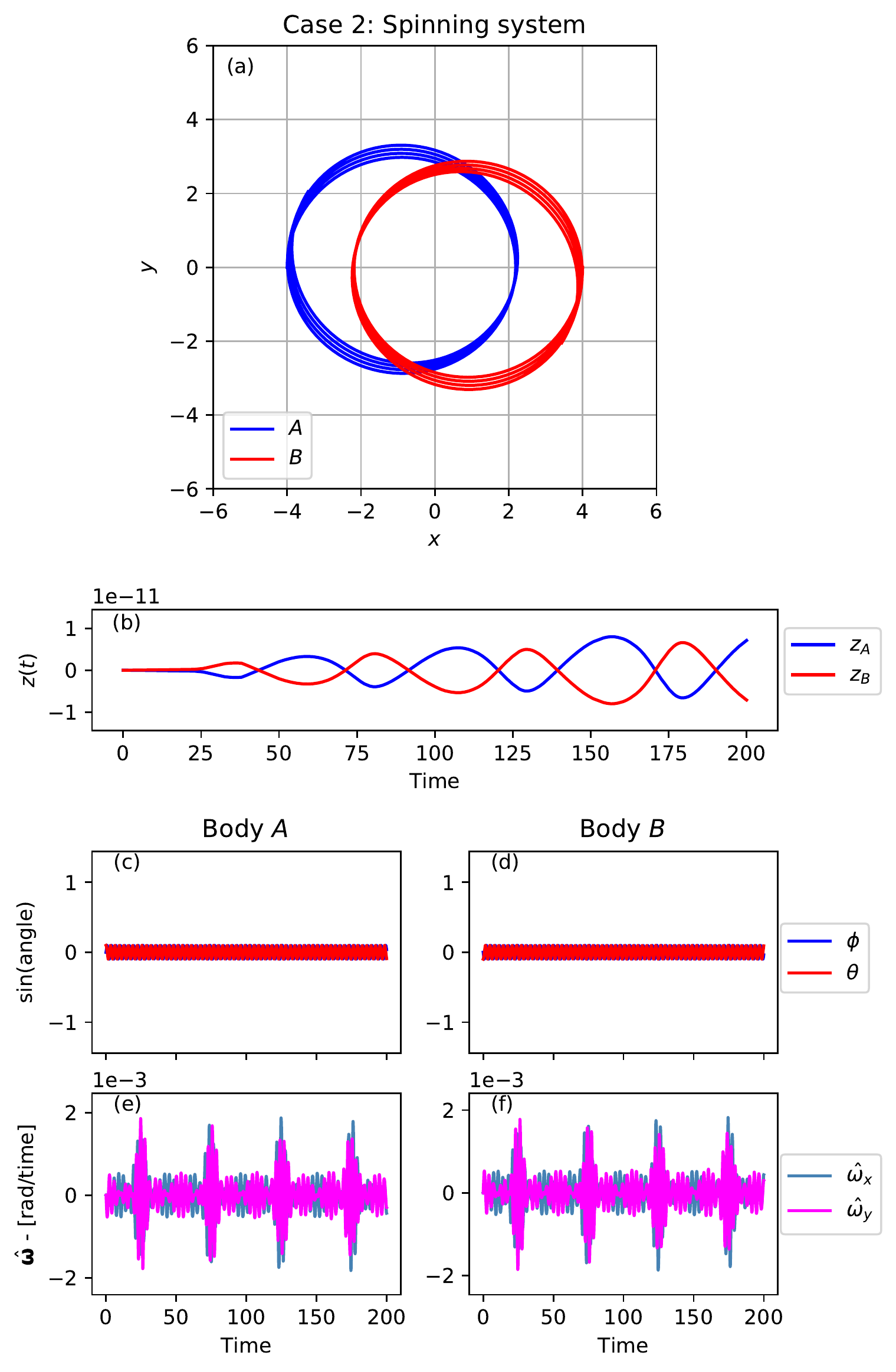}
\caption{Results of the translational and rotational motion of the spheroids for case 2. We have not shown $\hat{\omega}_z$ since it has a constant value of $\hat{\omega}_z = 2$.}
\label{fig:Case3Results}
\end{figure*}

Fig. \ref{fig:Case3Results}a shows the spheroid orbits projected onto the $xy$-plane. The orbits for this case closely resemble the case for two spheroids in co-planar motion \citep{2021CeMDA.133...27W}. Fig. \ref{fig:Case3Results}b shows their $z$-positions as a function of time. As in Case 1, a force along the $z$-axis can occur, but the motion in the $z$-direction is $11$ orders of magnitude smaller than the motion in the $x$ and $y$-directions for both spheroids.  

Figs. \ref{fig:Case3Results}c and \ref{fig:Case3Results}d show the rotation angles $\phi$ and $\theta$, plotted as sine functions, of bodies $A$ and $B$ respectively. As the spheroid spins around its $z$-axis, the angle $\psi$ increases linearly and will appear as a straight line with constant slope in the figures.  Meanwhile, the angles ($\phi$,$\theta$) oscillate between $-\pi/32$ and $\pi/32$. As $\psi$ is increasing linearly and ($\phi,\theta$) are both oscillating, the rotational motion of the two spheroids is very similar to that of a spinning top or a precessing gyroscope. Compared to case 1 in Sec. \ref{sec:tumbling_system}, where the spheroids tumbled, by spinning the spheroids around their corresponding axes of symmetry, the rotational motion stabilises. 

The angular velocity components $(\hat{\omega}_x,\hat{\omega}_y)$ are shown in Figs. \ref{fig:Case3Results}e and \ref{fig:Case3Results}f for bodies $A$ and $B$, respectively. Because of the rotational symmetry of the spheroid, $\hat{\omega}_z$ will remain constant at the value $\hat{\omega}_z = 2$ throughout the simulation, and is therefore excluded from the figures. The values of the $\hat{\omega}_x$ and $\hat{\omega}_y$ components peaks when the two spheroids are in close proximity. By increasing the initial angles by a factor of four, i.e. $\theta_0=\pi/8$, the amplitude of both the $\hat{\omega}_x$ and $\hat{\omega}_y$ components nearly doubles, indicating that the amplitude of these two angular velocity components is sensitive to initial angle.

\begin{figure}
\centering
\includegraphics[width=\linewidth]{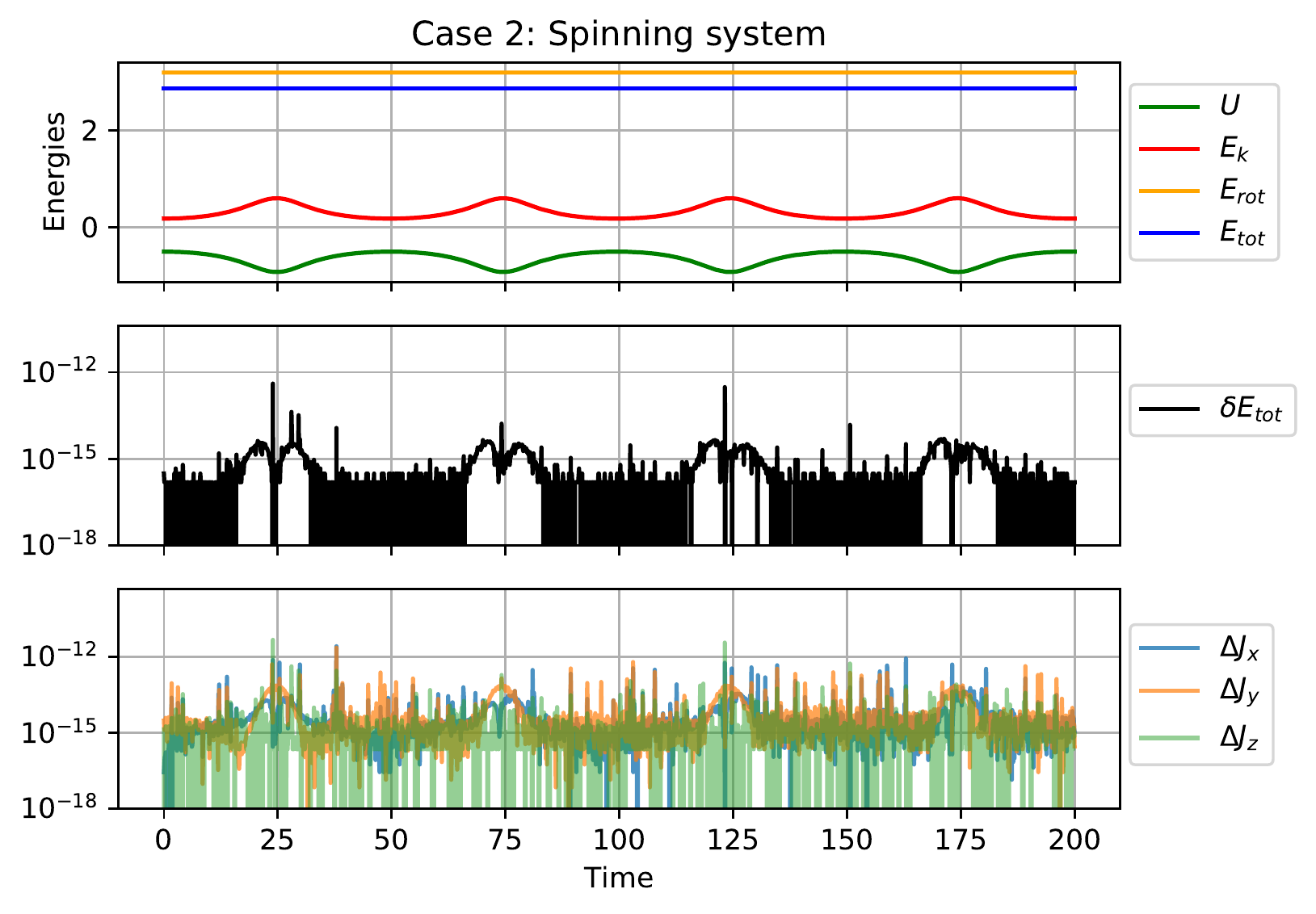}
\caption{Same as Fig. \ref{fig:Case2Energyxrot} but for case 2, where both spheroids have an initial angular velocity.
The total energy and total angular momentum of the system is conserved, as demonstrated by the small relative error in these two quantities.}
\label{fig:Spin_energy}
\end{figure}

The different parts of the total energy are shown in the top panel of Fig. \ref{fig:Spin_energy}. Due to the spheroids' rotations, the rotational energy now makes a significant contribution to the total energy, being greater than the kinetic energy from the  translational motion. However, because $\hat{\omega}_x$ and $\hat{\omega}_y$ are roughly 3 orders of magnitude smaller than $\hat{\omega_z}$, and $\hat{\omega}_z$ is constant, the rotational energy is nearly constant. Furthermore, both the total energy and total angular momentum remain constant in the simulation, as their respective errors are smaller than $10^{-12}$. 

\section{Application to (66391) 1999 KW4}
\label{sec:Result_real}
\begin{table}
\caption{The parameters and initial conditions used for the simulations. The orbital elements, obtained from \citet{2006Sci...314.1276O}, are those of Beta in orbit around Alpha. For the spheroid simulations, we set the semiaxes to be $a=b$.}
\label{table:Init_conditions_Moshup}
\renewcommand{\arraystretch}{1.2}
\begin{tabular}{lcc}
\hline\noalign{\smallskip}
Parameter & Alpha & Beta \\
\noalign{\smallskip}\hline
$a$, semiaxis along $x$ [km] & $0.766$ & $0.285$ \\
$b$, semiaxes along $y$ [km] & $0.748$ & $0.232$ \\
$c$, semiaxes along $z$ [km] & $0.674$ & $0.175$ \\
$m$, mass [$10^{12}$ kg] & $2.353$ & $0.135$ \\
$\phi_0$, initial rotation angle around $x$-axis [$^\circ$] & $3.2$ & $0.0$\\
$\hat{\omega}_{z,0}$, initial angular velocity [rad hr$^{-1}$] & $2.2728$ & $0.3611$ \\
\hline
$a_s$, semi-major axis [km] &  & $2.548$ \\
$e$, eccentricity & & $0.0004$ \\
$i$, inclination [$^\circ$] &  & $156.1$ \\
$\Omega$, longitude of ascending node [$^\circ$] &  & $105.4$ \\
$\omega$, argument of periapsis [$^\circ$] &  & $319.7$ \\
$M_0$, mean anomaly [$^\circ$]&  & $0.0$\\
\noalign{\smallskip}\hline
\end{tabular}
\end{table}

In this section, we apply our method to the asteroid binary system (66391) 1999 KW4, where we refer to the primary as Alpha, and to the secondary as Beta. We will study the dynamical evolution of the bodies and see how the outcome changes when the body shapes are changed from ellipsoids to spheroids by setting the semiaxes to be $a=b>c$. The simulations run with a timespan of 1 year. 

The physical parameters, as well as initial conditions, used in the simulations are shown in Tab. \ref{table:Init_conditions_Moshup}. We let Alpha start at rest from the origin of the inertial frame in the simulations. We also assume that Beta is located at the pericenter at $t=0$, and thus set the mean anomaly $M_0$ to be initially zero. \citet{2006Sci...314.1276O} also find that the angle between the rotation pole of Alpha and the binary orbit is between $0^\circ$ and $7.5^\circ$ with a nominal separation of $3.2^\circ$. As such, we let Alpha be initially rotated with $\phi_0 = 3.2^\circ$, whereas Beta initially remains non-rotated. It is assumed that the orbit of Beta is synchronous, and the initial angular velocity of Beta is set so that its rotation period is equal to its orbital period.

\begin{figure}
\centering
\includegraphics[width=\linewidth]{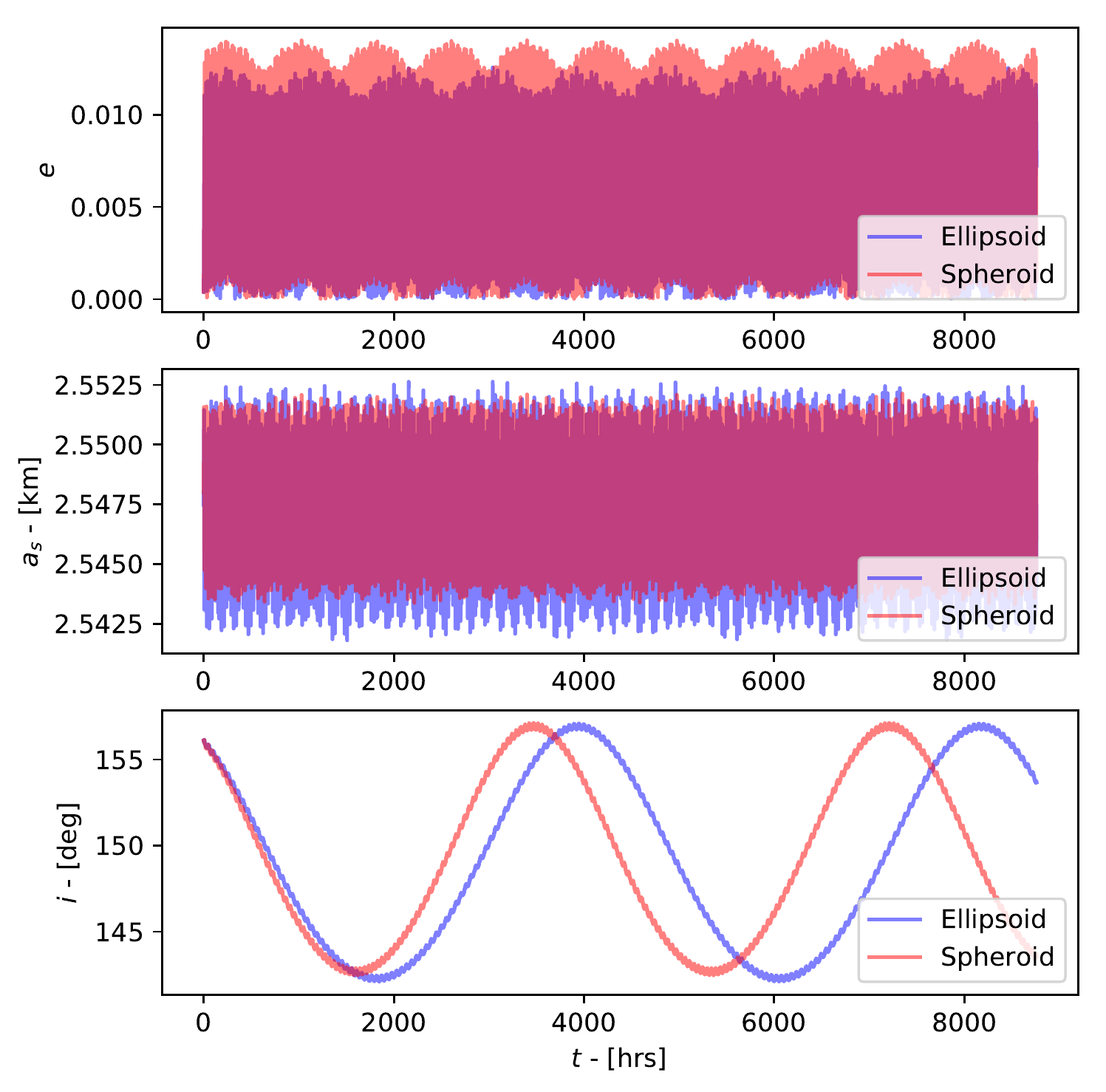}
\caption{Time evolution of the eccentricity, semi-major axis and inclination with respect to the original plane of orbit.}
\label{fig:KW1994_orbels}
\end{figure}

The top, middle and bottom panels of Fig. \ref{fig:KW1994_orbels} shows the eccentricity, semi-major axis and inclination of Beta respectively. The inclination is calculated with respect to the original plane of orbit. For a spheroidal model, the eccentricity can take slightly higher values, where the eccentricity peaks at $e=0.0140$ for the spheroidal model, while it peaks at $e=0.0126$ for the ellipsoidal model. This is lower to the eccentricity in the excited state of \citet{2008Icar..194..410F} \citep[see also][]{2014CeMDA.119..313C, 2017CeMDA.127..369H, 2017CeMDA.129..307S}, which find that the eccentricity can surpass $e=0.03$. 
The range of the semi-major axis are similar in both simulation types, where it takes values between $a_s\in[2.5418, 2.5526]$ km and averaging at $\bar{a}_s = 2.5471$ km for the ellipsoidal model, and $a_s \in [2.5433, 2.5521]$ km and averages at $\bar{a}_s = 2.5477$ km for the spheroidal model.

Because the orbital period is proportional to the semi-major axis, the fact that the average semi-major axis is larger in the spheroidal simulation, indicates that the orbital period is also longer. We find that the average orbital period of Squannit is $\bar{T} = 17.4116$ hrs for the ellipsoidal simulation and $\bar{T}=17.4177$ for the spheroidal simulation. The period of the inclination is longer for the ellipsoidal model, where the period is approximately $3900$ hrs between the two maxima, whereas the period is approximately $3400$ hrs for the spheroidal model.

\begin{figure}
\centering
\includegraphics[width=\linewidth]{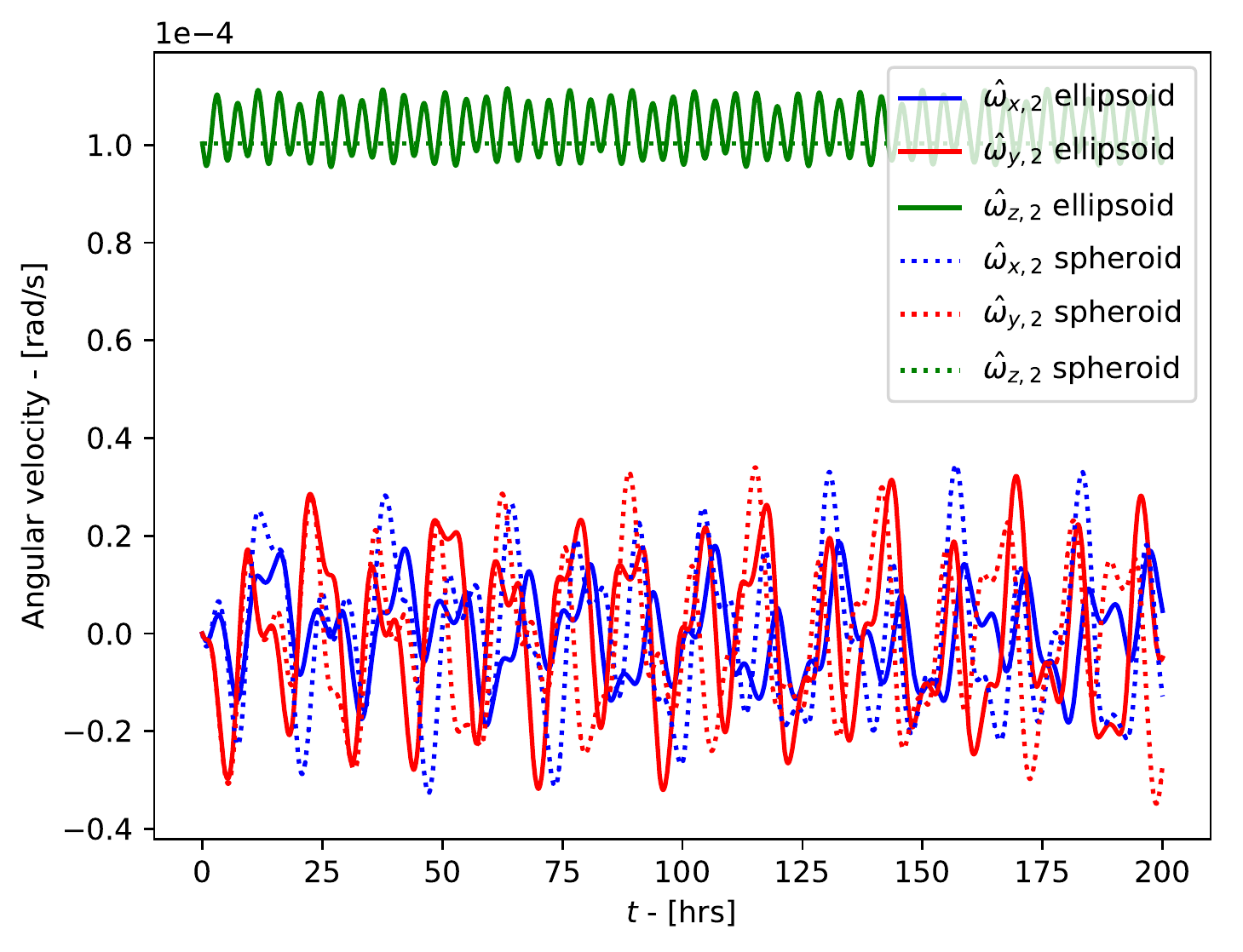}
\caption{The angular velocity components of Beta as a function of time. Solid lines corresponds to the bodies take ellipsoidal shapes, while the dotted lines correspond to the bodies with spheroidal shapes.}
\label{fig:KW1994_angvel_Second}
\end{figure}

Fig. \ref{fig:KW1994_angvel_Second} shows the angular velocity components of the secondary for the first $200$ hours of the simulation. Here, we find that the range of $\hat{\omega}_z$ is $\hat{\omega}_z \in [9.55, 11.16]\cdot 10^{-5}$ rad/s. Compared to the findings of \citet{2008Icar..194..410F}, the range range is smaller than the excited state, but also larger than the relaxed state, than that of Fahnestock and Scheeres. Another major difference is in the components of both $\hat{\omega}_x$ and $\hat{\omega}_y$. The work of Fahnestock and Scheeres find that both of these components change very insignificantly for both the excited and relaxed configurations. Our findings, however, show that the components respectively oscillates around the values $\hat{\omega}_x \in [-1.89, 1.86]\cdot 10^{-5}$ rad/s, $\hat{\omega}_y \in [-3.20, 3.23]\cdot 10^{-5}$ rad/s. As previously mentioned in dimensionless test scenario ``spinning system'' (see Sec. \ref{sec:Spinning_system}), the amplitude of the $\hat{\omega}_x$ and $\hat{\omega}_y$ components could be affected by the initial angle. The difference in our result, and the one of Fahnestock and Scheeres, could  therefore be due to the difference in the initial angles.

Changing the body shapes from ellipsoids to spheroids significantly affects $\hat{\omega}_z$. As seen in Fig. \ref{fig:KW1994_angvel_Second}, by allowing the bodies to take a spheroidal shapes, $\hat{\omega}_z$ for Beta becomes constant as opposed to oscillating when it had an ellipsoidal shape. This is because spheroids are rotationally symmetric, and no torques can act to change the angular velocity in the $z$-direction. This is also what was seen in the dimensionless test scenarios in Sec. \ref{sec:Results_test}. The $\hat{\omega}_x$ and $\hat{\omega}_y$ components, on the other hand, still oscillates between the values seen for an ellipsoidal shape. The angular velocity of Beta in the spheroidal case is similar to the result that was previously shown in Fig. \ref{fig:Case3Results}e and \ref{fig:Case3Results}f, in which $\hat{\omega}_z$ was constant and both $\hat{\omega}_x$ and $\hat{\omega}_y$ were oscillating over time.

\begin{figure}
\centering
\includegraphics[width=\linewidth]{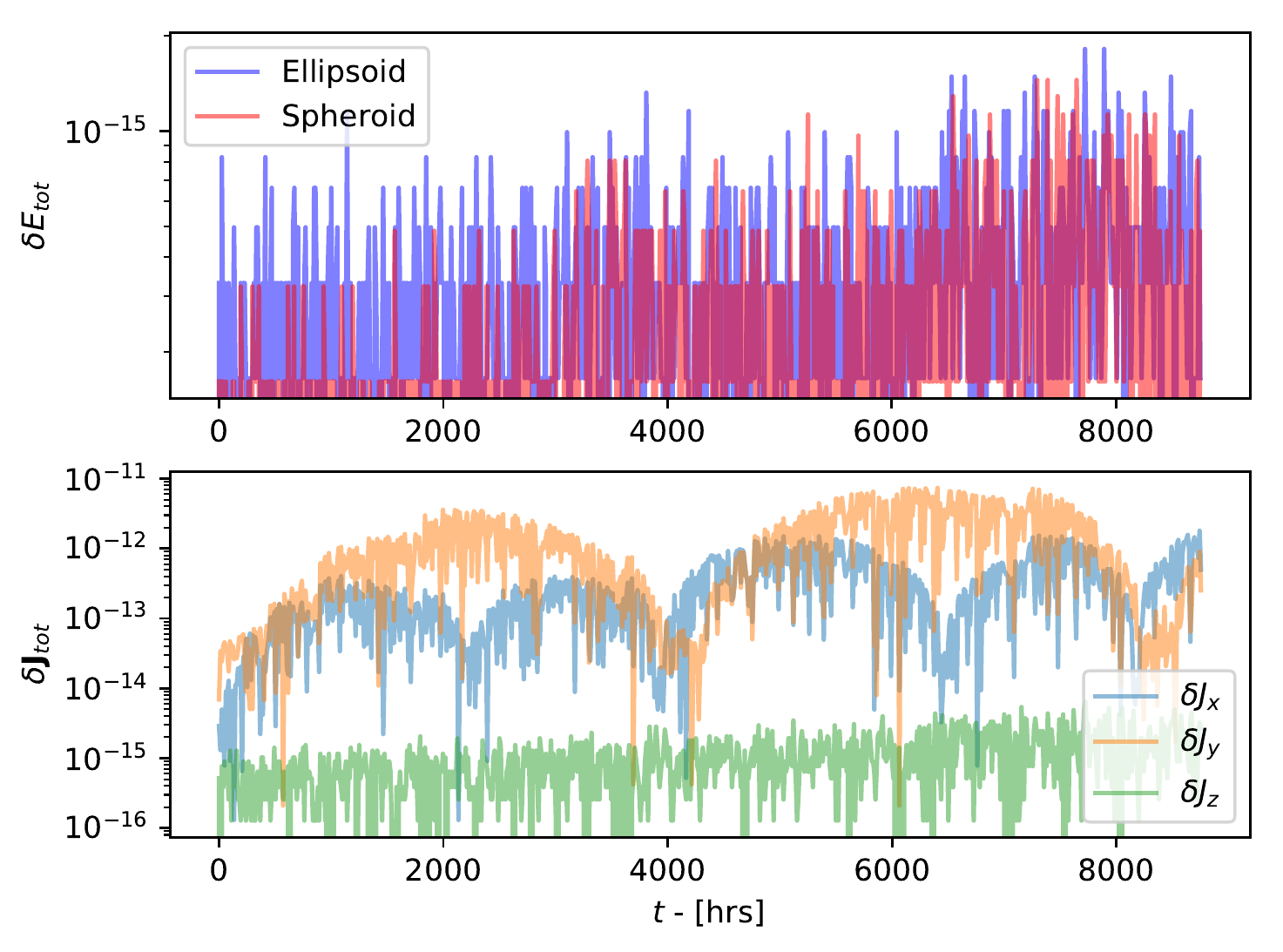}
\caption{\textit{Top}: The relative error of the total energy of the (66391) 1999 KW4 system, where the blue and red curves correspond to the ellipsoidal and spheroidal simulations respectively. \textit{Bottom}: The relative error of the components of the total angular momentum for the ellipsoidal simulation.}
\label{fig:KW1994_energy_error}
\end{figure}

Fig. \ref{fig:KW1994_energy_error} shows the relative error in the total energy and the relative error of the components of the total angular momentum in the top and bottom panels respectively. The blue and red curves correspond to simulations where the bodies take ellipsoidal and spheroidal shapes. The relative error of the total energy is smaller than $10^{-14}$ for both the ellipsoidal and spheroidal simulations. For the total angular momentum, we only show the relative error in the ellipsoidal simulation. Here, the relative errors in each component are smaller than $10^{-11}$. The errors in the spheroidal simulation are similar to the ellipsoidal simulation. Although, throughout the simulation, there is a drift in the total angular momentum, causing the errors in each component to increase over time. 

\section{Summary and discussion}
\label{sec:summary}
\begin{table}
\caption{CPU time used for each simulation, as well as simulation time, presented in the paper. Note that the simulation time for Case 1 and Case 2 have a dimensionless quantity.}
\label{table:sim_time}
\begin{tabular}{lcc}
\hline
Scenario & CPU time & Simulation time   \\
\hline
Case 1 & 22.8 s & 200\\
Case 2 & 129.0 s & 200\\
1999 KW4, ellipsoids & 4.6 hrs & 1 yr \\
1999 KW4, spheroids & 1.1 hrs & 1 yr \\
\hline
\end{tabular}
\end{table}
We have validated and explored an alternative method for simulating the dynamics of a fully three dimensional rigid two-body problem suggested by \citet{Conway_2016}. The method, which is based on vector potentials instead of scalar potentials, uses surface integrals to determine the force, torque and mutual potential energy between two bodies. \citet{2021CeMDA.133...27W} outlined the surface integration method in detail, and tested it for a pair of coplanar spheroids and thin disks. In this work, we extend the work and apply the method to pairs of ellipsoids and spheroids that can be randomly oriented with respect to each other, hence torques and angular momentum exchange is included. 

Tab. \ref{table:sim_time} shows a summary on the CPU time used for each simulation. The CPU time used for the dimensionless simulations varied between 20 seconds to two minutes. For the (66391) 1999 KW4 system, the CPU time required varied between one hour to four hours. 

Two dimensionless cases were studied, where both test scenarios considered spheroidal body shapes. In the first case, both the spheroids are initially rotated around their body-fixed $x$-axes. Despite the lack of initial motion along the $z$-direction, the initial rotational tilt allowed a small force component along the $z$-direction to take place. This results to a small motion along the $z$-direction over time, although the motion is 11 orders of magnitude smaller than the motion along the $x$ and $y$ direction. Furthermore, the rotational motion of the spheroids develops into tumbling motion.
The second dimensionless case considered is similar to the first one, where both bodies are now initially rotated around their body-fixed $y$-axes. However, in this scenario, both bodies also start with an angular velocity about their body-fixed $z$-axes. The motions of the spheroids' in this case closely resembled two tops spinning in orbit around the common center of mass. By spinning the spheroids about their axes of symmetry, the rotational motion stabilises so that it is no longer tumble-like.

Finally, we apply the method on the asteroid binary system (66391) 1994 KW4. In this scenario, we consider two types of simulations: one where both bodies have ellipsoidal shapes and one where both have spheroidal shapes. 

We compare the difference in the dynamical evolution of Beta when the bodies had ellipsoidal and spheroidal shapes. The eccentricity, on average, is larger in the spheroidal simulation. Furthermore, the values of the eccentricity is smaller than the findings of of \citet{2008Icar..194..410F}. The values of the semi-major axis are similar for both simulation types, but the average semi-major axis is slightly larger for the spheroidal simulation compared to the ellipsoidal simulation. This also indicates that the orbital period becomes longer when the bodies take ellipsoidal shapes. We also find that the time period it would take for the inclination to reach its maximum are longer when both bodies took ellipsoidal shapes.

The angular velocity components of Beta is also studied. The results are also compared to the findings of \citet{2008Icar..194..410F}, where we find that the $\hat{\omega}_z$ component, for the ellipsoidal simulation, is similar to the findings of Fahnestock and Scheeres. However, the evolution of both $\hat{\omega}_x$ and $\hat{\omega}_y$ are different, in which we find that these components are oscillating with larger amplitudes compared to the findings of Fahnestock and Scheeres, which is due to the difference in the initial conditions.

Studying the errors in the total energy and total angular momentum serves as a check of simulation accuracy. We find that the errors for both the total energy and total angular momentum are smaller than $10^{-12}$ for all simulations presented. The errors, which are numerical in origin, are small enough to demonstrate that our model conserves energy and angular momentum.

While our method has only been demonstrated here for a handful of scenarios, the method can also be generalized to an $N$-body simulation, which can be used to simulate e.g. an asteroid triple system and even include the gravitational pull from the planets in the Solar system.

\begin{acknowledgements}
The authors thank the anonymous referee for the helpful comments and suggestions that improved the manuscript.
\end{acknowledgements}

\begin{appendix}
\section{Gravitational field of an ellipsoid and spheroid}
\label{sec:Appendix_gravfield}
The gravitational field is required to compue the mutual potential energy in Eq. \eqref{EqnA3_energy}. We will here derive an analytical expression of $\mathbf{g} = (g_x, g_y, g_z) = \nabla \Phi$ based on the expression in Eqs. \eqref{eq:Potential_elliptic} and \eqref{eq:Oblate_spheroid}. It should be noted that, while $\kappa$ is a function of $(x,y,z)$, when taking the partial derivatives of the gravitational potential $\Phi$, $\kappa$ can be treated as a constant \citep{MacMillan1930}.

For a general ellipsoid, the components of the gravitational field thus become
\begin{align}
g_x &= \frac{4x\pi\rho abc}{\sqrt{a^2-c^2}}\frac{E(\omega_\kappa) - F(\omega_\kappa)}{a^2-b^2} \\
g_y &= \frac{4y\pi\rho abc}{\sqrt{a^2-c^2}}\Bigg[\frac{F(\omega_\kappa)}{a^2-b^2} - \frac{(a^2-c^2)E(\omega_\kappa)}{(a^2-b^2)(b^2-c^2)} \nonumber \\
&+ \frac{(c^2+\kappa)}{b^2-c^2}\frac{\sqrt{a^2-c^2}}{\sqrt{(a^2+\kappa)(b^2+\kappa)(c^2+\kappa)}}\Bigg]  \\
g_z &= \frac{4z\pi\rho abc}{\sqrt{a^2-c^2}}\Bigg[\frac{E(\omega_\kappa)}{b^2-c^2} - \frac{(b^2+\kappa)}{b^2-c^2}\frac{\sqrt{a^2-c^2}}{\sqrt{(a^2+\kappa)(b^2+\kappa)(c^2+\kappa)}}\Bigg].
\end{align}
For an oblate spheroid, the components of $\mathbf{g}$ are
\begin{align}
g_x &=\frac{2\pi\rho x a^2 c}{a^2-c^2}\left[\frac{\sqrt{c^2 + \kappa}}{a^2 + \kappa} - \frac{1}{\sqrt{a^2-c^2}}\sin^{-1}\left(\sqrt{\frac{a^2-c^2}{a^2+\kappa}}\right)\right] \\
g_y &=\frac{2\pi\rho y a^2 c}{a^2-c^2}\left[\frac{\sqrt{c^2 + \kappa}}{a^2 + \kappa} - \frac{1}{\sqrt{a^2-c^2}}\sin^{-1}\left(\sqrt{\frac{a^2-c^2}{a^2+\kappa}}\right)\right] \\
g_z &= \frac{4\pi\rho z a^2 c}{a^2-c^2}\left[\frac{1}{\sqrt{a^2-c^2}}\sin^{-1}\left(\sqrt{\frac{a^2-c^2}{a^2+\kappa}}\right) - \frac{1}{\sqrt{c^2+\kappa}}\right].
\end{align}
The value of $\kappa$, for both the ellipsoid and spheroid cases, still satisfies Eq. \eqref{eq:kappa_eq}.
\end{appendix}

\bibliographystyle{spbasic}
\bibliography{references} 
\end{document}